# Enhanced Superconductivity at Quantum-Critical KTaO$_3$ Interfaces

Jieun Kim[1,†], Muqing Yu[2,†], Ahmed Omran[2], Jiangfeng Yang[1], Ranjani Ramachandran[2], William O. Nachlas[3], Patrick Irvin[2], Jeremy Levy[2,*] and Chang-Beom Eom[1,*]

[1]Department of Materials Science and Engineering, University of Wisconsin-Madison, Madison, 53706, Wisconsin, USA.

[2]Department of Physics and Astronomy, University of Pittsburgh, Pittsburgh, 15260, Pennsylvania, USA.

[3]Department of Geoscience, University of Wisconsin-Madison, Madison, 53706, Wisconsin, USA.

†These authors contributed equally to this work.

*Corresponding author(s). E-mail(s): jlevy@pitt.edu; ceom@wisc.edu;

**Abstract**

Superconductivity at oxide interfaces has intrigued researchers for decades, yet the underlying pairing mechanism remains elusive. Here we demonstrate that proximity to a ferroelectric quantum critical point dramatically enhances interfacial superconductivity in KTaO$_3$. By precisely tuning KTaO$_3$ to its quantum critical composition through 0.8% niobium doping, we achieve a near-doubling of the superconducting transition temperature, reaching 2.9 K. Remarkably, a dome-shaped carrier density dependence emerges exclusively at the quantum critical point, contrasting sharply with the linear scaling observed in undoped interfaces. Our findings establish ferroelectric quantum criticality as a powerful mechanism for enhancing superconductivity and provide compelling evidence for soft-phonon-mediated pairing in these systems.

**Introduction**

Oxide interfaces represent a frontier in condensed matter physics, hosting emergent phenomena absent in their bulk counterparts[1–3]. Among these systems, two-dimensional electron gases at the interface of insulating oxides, such as $LaAlO_3/SrTiO_3$ and $LaAlO_3/KTaO_3$, have been a subject of intense research due to their surprising metallic and superconducting properties. The existence of superconductivity in these systems is puzzling, given their extremely low carrier densities and exceptionally low superconducting transition temperatures ($T_c$).

A particularly intriguing puzzle arises when comparing the magnitudes of $T_c$ in $SrTiO_3$ and $KTaO_3$. While both are incipient ferroelectrics $KTaO_3$ interfaces achieve higher superconducting transition temperatures than $SrTiO_3$ interfaces[1,3], despite $SrTiO_3$ being closer to ferroelectric instability[4,5]. Furthermore, $KTaO_3$-based interfaces exhibit superconductivity, only at specific crystallographic orientations[3,6–11] and shows anomalous carrier density scaling[6,10]. These paradoxes suggest a fundamental gap in our understanding of the pairing mechanism. While quantum criticality drives superconductivity by providing the pairing glue for electrons in unconventional superconductors[12–16], its role in oxide interfaces remains largely unexplored.

Recent theoretical work by Edge et al.[17] provides a framework to resolve this paradox[17]. Their model suggests that the superconducting dome observed in $SrTiO_3$ originates from a competition between the emergence of a Fermi surface and the carrier-induced suppression of ferroelectric quantum fluctuations. $KTaO_3$ offers an ideal platform to investigate this connection: quantum fluctuations suppress its expected ferroelectric transition, maintaining cubic symmetry to absolute zero[18]. This places $KTaO_3$ in the ideal regime to test the hypothesis that ferroelectric quantum criticality can enhance pairing through soft phonon fluctuations. Crucially, Nb substitution tunes $KTa_{1-x}Nb_xO_3$ precisely to a ferroelectric quantum critical point at a well-defined, low Nb content (x = 0.8%[19]). At this composition, quantum fluctuations, and therefore any phonon-mediated enhancement of pairing, are expected to maximize.

Here we systematically explore how proximity to the quantum critical point governs interfacial superconductivity in Nb-doped $KTaO_3$, revealing a profound enhancement mechanism linked to soft phonon dynamics. To isolate quantum criticality effects from disorder effects, we combine electrostatic gating with adsorption-controlled epitaxial growth with precise control of K stoichiometry and Nb content. Electronic grade Nb-doped $KTaO_3$ films and atomically clean $LaAlO_3/KTa_{1-x}Nb_xO_3$ interfaces reduce disorder and stabilize the (111) orientation where superconductivity is most robust[20]. This platform allows quantitative links to be established between transport, carrier density, and soft mode tuning, and it separates intrinsic quantum critical responses from extrinsic effects that often mask narrow composition windows.

**Results**

**Quantum critical enhancement of superconductivity**

We performed transport measurements to examine how Nb doping affects interfacial superconductivity by growing epitaxial $KTa_{1-x}Nb_xO_3$ thin films with four precisely controlled compositions spanning the quantum critical point: x = 0, 0.4%, 0.8%, and 1.6%. All films are grown on (111)-oriented $KTaO_3$ substrates with amorphous $LaAlO_3$ overlayers to create conducting interfaces using a recently developed hybrid pulsed laser deposition technique[20] (see Methods). The temperature dependence of the sheet resistance ($R_{sq}$) of the films, measured over the entire surface of the films in a Van der Pauw geometry, shows metallic behavior for all films, demonstrating high-quality epitaxial growth of the $KTa_{1-x}Nb_xO_3$ films with precise K stoichiometry[20]. To isolate quantum criticality effects from sheet carrier density ($n_{2D}$) variations, we select samples with comparable $n_{2D}$ values at low temperatures (Fig. 1b). The Nb-doped samples have $n_{2D}$ = 5.6-6.0×$10^{13}$ cm$^{-2}$ at 4 K, while the undoped sample has $n_{2D}$ = 8.11 × $10^{13}$ cm$^{-2}$. We patterned Hall bars along the [11-2] and determined the superconducting transition temperature $T_c$ at 50% of the normal state $R_{sq}$, just above the transition. The $R_{sq}$ drops to zero for all samples, attesting to the quality of the $LaAlO_3/KTa_{1-x}Nb_xO_3$ (111) interfaces with minimal growth-induced disorder and to the resultant uniformity of the superconducting phase. Remarkably, the Nb-doped samples exhibit enhanced $T_c$ (2.05-2.5 K) compared to the undoped sample (1.53 K), despite their lower $n_{2D}$ (Fig. 1c). Both $T_c$ and critical current density ($J_c$) peak precisely at x = 0.8% (insets of Fig. 1c,d), corresponding to the known bulk ferroelectric quantum critical point[19]. This striking correlation demonstrates that quantum criticality, not carrier density, drives the enhancement.

**Emergence of a superconducting dome**

Having established enhancement of $T_c$ at the quantum critical point, we explore how carrier density affects this phenomenon in $LaAlO_3/KTa_{0.992}Nb_{0.008}O_3$ (111) interfaces. By varying the $LaAlO_3$ growth temperature ($T_g$) from 100 to 500 °C, we can systematically tune $n_{2D}$ of $LaAlO_3/KTa_{0.992}Nb_{0.008}O_3$ (111) from 4.67 to 7.48 x $10^{13}$ cm$^{-12}$. The resulting distributions of $n_{2D}$ and mobility ($\mu$) at 10 K (Fig. 2a) show that the mobility distribution is shifted to lower values in the Nb-doped samples compared to the undoped samples. This shift is consistent with additional impurity scattering introduced by dilute Nb dopants but it evolves smoothly with $n_{2D}$ and does not produce a peak that could mimic the non-monotonic $T_c$. Figure 2b reveals a non-monotonic dependence of $T_c$ on $n_{2D}$, with a maximum of 2.9 K at $n_{2D}$ = 6.35 × $10^{13}$ cm$^{-2}$ - nearly double the value for undoped interfaces. Systematic measurements on 20 samples reveal a striking dome-shaped dependence at the quantum critical point (Fig. 2c), contrasting sharply with the linear relationship in undoped $KTaO_3$[7].

**Electrostatic gate tuning at the quantum critical composition**

The superconducting dome in $LaAlO_3/KTa_{0.992}Nb_{0.008}O_3$ (111) interfaces is confirmed not only by varying the $n_{2D}$ using different $LaAlO_3$ growth temperatures, but also by directly modulating the $n_{2D}$ at the quantum critical composition. Due to the relatively low $\delta n_{2D}$ (< 2 x $10^{13}$ cm$^{-2}$) that can be induced by electrostatic gating in $KTaO_3$ and the large range of $n_{2D}$ where $KTaO_3$ is superconducting ($\approx 10^{14}$ cm$^{-2}$), we chose three samples prepared with $T_g$ = 250, 475, 550 °C, which represent $LaAlO_3/KTa_{0.992}Nb_{0.008}O_3$ (111) interfaces with carrier densities in the under-doped, optimally-doped, and over-doped regimes, respectively. Fig. 3a shows the $R_{sq}$ (T) along the [11-2] for the 475 °C sample. At $V_G$ = 0 V, the superconducting transition occurs at $T_c$ = 2.9 K. Sweeping $V_G$ from -26 to 120 V shifts the entire transition while the ground state remains at zero resistance, indicating a strong $V_G$ dependence of $T_c$. Fig. 3b shows the modulation of $n_{2D}$ and $\mu$ for the 250, 475, 550 °C samples. For all samples, $n_{2D}$ and $\mu$ decrease monotonically as $V_G$ is lowered from 50 V to -50 V, as expected from enhanced effective disorder near the interface. In contrast to the monotonic decrease in $n_{2D}$, we observe contrasting evolution of $T_c$ with $V_G$ (Fig. 3c). In the 250 °C sample, $T_c$ shows nearly monotonic decrease as $V_G$ is lowered whereas in the 475 °C sample $T_c$ initially increases to a peak of 3.0 K before decreasing as $n_{2D}$ passes the top of the superconducting dome at $n_{2D}$ = 5.39 x $10^{13}$ cm$^{-2}$. In the 550 °C sample, $T_c$ exhibits a strongly inverse correlation with $n_{2D}$, as expected in the over-doped regime. Taken together, the 250, 475, 550 °C samples map a nearly continuous superconducting dome in the $T_c$-$n_{2D}$ diagram. The 250 °C and 550 °C samples occupy the low and high $n_{2D}$ sides and trace the rising and falling branches, respectively, while the 475 °C samples connect the two branches, with the apex located near $n_{2D} = 5.39 \times 10^{13}$ cm$^{-2}$ and $T_c = 3.0$ K. Over the same $n_{2D}$ range, the undoped KTaO$_3$ reference increases approximately linearly with and shows no maximum. Collectively, these results establish a non-monotonic dependence of $T_c$ on $n_{2D}$ at the ferroelectric quantum critical point.

**Discussion**

The substantial enhancement of superconductivity at the ferroelectric quantum critical point reveals a fundamental connection between quantum lattice fluctuations and Cooper pair formation in LaAlO$_3$/KTa$_{0.992}$Nb$_{0.008}$O$_3$ (111) interfaces. Unlike conventional phonons, whose occupation is thermally suppressed at low temperatures, quantum critical polar fluctuations persist down to absolute zero. These fluctuations provide an unusually robust pairing channel, explaining why $T_c$ nearly doubles despite reduced carrier density relative to undoped KTaO$_3$ interfaces. Our results therefore provide compelling evidence for a long-standing hypothesis: the soft-phonon-mediated pairing can dominate in oxide superconductors when a ferroelectric mode is tuned to criticality.

The emergence of a dome-shaped dependence of $T_c$ on $n_{2D}$ reinforces this mechanism. In undoped KTaO$_3$, $T_c$ scales linearly with $n_{2D}$, consistent with conventional BCS behavior where a growing Fermi surface expands the phase space for electron-phonon interaction. At the quantum critical composition, however, this linearity breaks down and $T_c$ first rises at low $n_{2D}$ as polar fluctuations strongly enhance pairing, but at higher $n_{2D}$ these fluctuations are progressively screened by mobile carriers, resulting in an optimal density and a dome-shaped profile in the $T_c$-$n_{2D}$ diagram. This non-monotonicity mirrors that observed in doped SrTiO$_3$ and in a wide range of unconventional superconductors near nematic or magnetic quantum critical points, highlighting the universality of quantum criticality as a driver of superconductivity[15,17,21]. We note that the emergence of a superconducting dome at the quantum critical point in the regime of linear dependence of $T_c$ on $n_{2D}$ leads to highly anisotropic enhancement of $T_c$ on either side of the dome. On the over-doped side, $T_c$ collapses rapidly to that of the undoped baseline within $\delta n_{2D} \approx 1.8 \times 10^{13}$ cm$^{-2}$ beyond the peak, whereas the under-doped side exhibits a three-fold enhancement of $T_c$ even at $\delta n_{2D} \approx 2.1 \times 10^{13}$ cm$^{-2}$. These quantitative comparisons demonstrate that quantum critical polar fluctuations enhance superconductivity over a broad under-doped range, whereas their carrier-induced screening at high $n_{2D}$ sharply suppresses the enhancement. This asymmetry distinguishes the superconducting dome at the ferroelectric quantum critical point in KTaO$_3$ from the more symmetric superconducting domes found in SrTiO$_3$, cuprates, and iron pnictides near their quantum criticalities[15,22,23], underpinning the pronounced role of polar fluctuations in this system. At the same time, the observation that $T_c$ peaks precisely at the ferroelectric quantum critical point establishes the general framework for soft-phonon-mediated superconductivity put forward by earlier theoretical[17] and experimental works, opening the door to engineering similar enhancements of superconductivity in other quantum paraelectric and correlated systems near structural or magnetic quantum critical points.

Future experiments could directly probe soft phonon spectra at these interfaces using inelastic neutron scattering or ultrafast spectroscopy, which will be essential for quantifying the dynamics of the critical mode and its coupling to the superconducting condensate[24–29]. The interplay between strong Rashba spin-orbit coupling in $KTaO_3$[30] and quantum critical superconductivity may enable exotic phases including topological superconductivity. Finally, our ability to position electronic-grade $KTaO_3$ precisely at its ferroelectric quantum critical point by Nb doping underlines the promise of advanced synthesis approaches. Continued progress in adsorption-controlled epitaxy and related techniques[20,31,32] will expand the accessible phase space for volatile-refractory systems and correlated oxides, enabling systematic exploration of structural, magnetic, and electronic quantum criticalities. The nearly two-fold enhancement of $T_c$ achieved here illustrates how synthesis-driven access to critical regimes can serve as a powerful lever for engineering superconductivity, and bodes well for future advances in the design of next-generation quantum materials with tunable and enhanced functionalities.

## Methods

### Film synthesis and structural characterization

Epitaxial $KTa_{1-x}Nb_xO_3$ thin films were grown using hybrid pulsed laser deposition (PLD) combining a KrF excimer laser (248 nm) with an effusion cell for potassium delivery. Films were deposited on (111)-oriented $KTaO_3$ substrates at 973 K in $10^{-6}$ Torr. Ceramic targets of $(1-x)Ta_2O_5-xNb_2O_5$ (x = 0, 0.004, 0.008, 0.016) were ablated at 0.5 J/cm$^2$ and 20 Hz. Potassium was supplied from $K_2O$ powder heated to 750 K in a tantalum-shielded MgO crucible. Growth rate was approximately 2.5 unit cells/min. Amorphous $LaAlO_3$ overlayers (3 nm) were deposited in situ at 673 K in $10^{-5}$ Torr using conventional PLD from a single crystal target. Structural quality was confirmed by X-ray diffraction and atomic force microscopy (Supplementary Fig. 1).

### Compositional analysis

Nb concentrations were verified using field-emission electron probe microanalysis (FE-EPMA) with wavelength-dispersive spectroscopy. The Nb L$\alpha$ peak position was calibrated using $LiNbO_3$ reference standards. Measurements employed 10 kV accelerating voltage, 400 nA beam current, and 20 μm spot size. X-ray intensities at the Nb L$\alpha$ position increased systematically with nominal doping concentration (Supplementary Fig. 2).

### Transport measurements

Electrical transport was measured using van der Pauw geometry from 2-300 K. Sheet resistance and Hall measurements employed DC current reversal to eliminate thermoelectric voltages. Carrier density was extracted from Hall slopes: $n_{2D} = 1/[(dR_H/dB)e]$. Superconductivity measurements below 2 K utilized a dilution refrigerator with four-terminal configuration. Current-voltage characteristics employed current biasing through 300 kΩ series resistance. Zero-bias resistance used lock-in detection (10 nA, 13 Hz). High-field magnetoresistance up to 18 T was measured in a separate dilution system (Supplementary Fig. 3).

**Device fabrication**

Hall bars were patterned using photolithography and reactive ion etching. Gold alignment markers were first deposited. AZ4210 photoresist protected Hall bar regions during 18-minute $BCl_3/Cl_2/Ar$ plasma etching (100 W). Etching depth of 50 nm was confirmed by atomic force microscopy.

**Supplementary information.** Supplementary Information accompanies this paper.

**Acknowledgements.** Thin film synthesis at the University of Wisconsin–Madison was supported by the US Department of Energy (DOE), Office of Science, Office of Basic Energy Sciences (BES), under award number DE-FG02-06ER46327. CBE acknowledges support for this research through the Gordon and Betty Moore Foundation's EPiQS Initiative, Grant GBMF9065 and a Vannevar Bush Faculty Fellowship (ONR N00014-20-1-2844). J.L. acknowledges NSF (DMR-2225888). Both C.B.E. and J.L. acknowledge ONR MURI (N00014-21-1-2537). We acknowledge helpful discussions with F. Yang and L. Q. Chen.

**Author contributions.** J.K. and C.B.E. conceived the project. C.B.E. and J.L. supervised the project. J.K. and J.Y. fabricated and characterized thin films. J.K., M.Y., A.O., and R.R. performed transport measurements. M.Y., A.O., and R.R. performed superconductivity measurements. W.N. performed compositional analysis. J.K., M.Y., J.L., and C.B.E. wrote the manuscript. All authors discussed results and commented on the manuscript.

**Competing interests.** The authors declare no competing interests.

**Data availability.** The data supporting the findings of this study are available from the corresponding authors upon reasonable request.

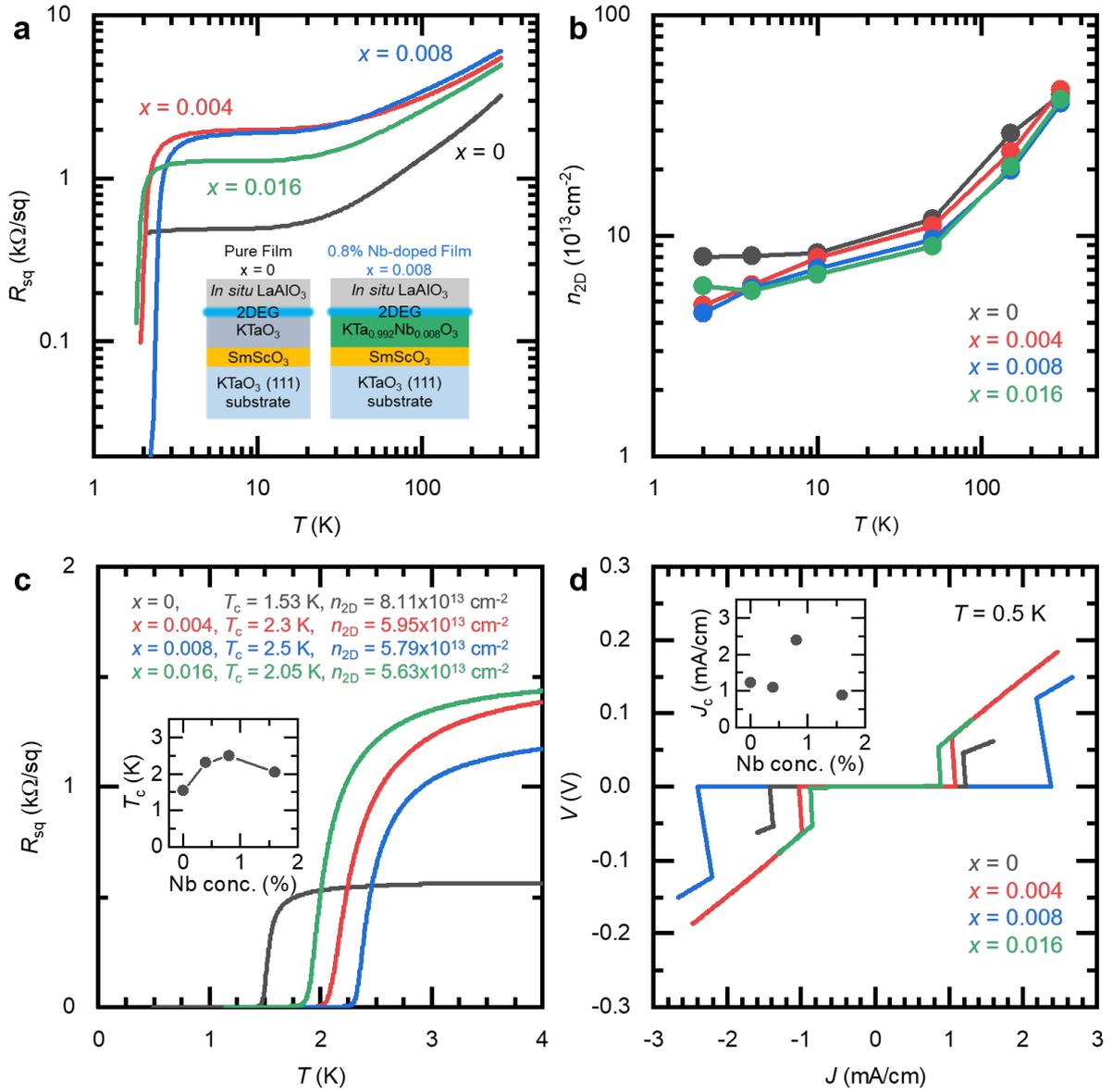

**Figure 1 | Quantum critical enhancement of superconductivity. a**, Temperature-dependent sheet resistance for different Nb concentrations. **b**, Carrier density remains comparable across samples. **c**, Superconducting transitions reveal maximum $T_c$ at the quantum critical composition (x = 0.8%). Inset: $T_c$ versus doping concentration. **d**, Critical current density also maximizes at x = 0.8%. Inset: Critical current versus doping.

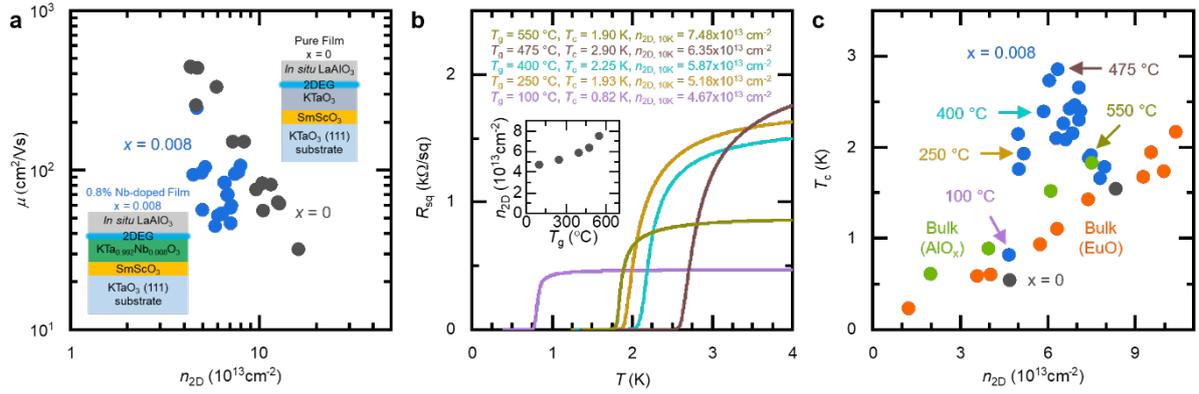

**Figure 2 | Superconducting dome at the quantum critical point. a**, Mobility and carrier density distributions for undoped (black) and quantum-critical (blue) interfaces. **b**, Superconducting transitions at different carrier densities. Maximum $T_c$ of 2.9 K occurs at intermediate density. **c**, Systematic mapping reveals dome-shaped $T_c$ dependence, contrasting with linear scaling in undoped samples. Bulk (EuO) is from C. Liu et al.[7]

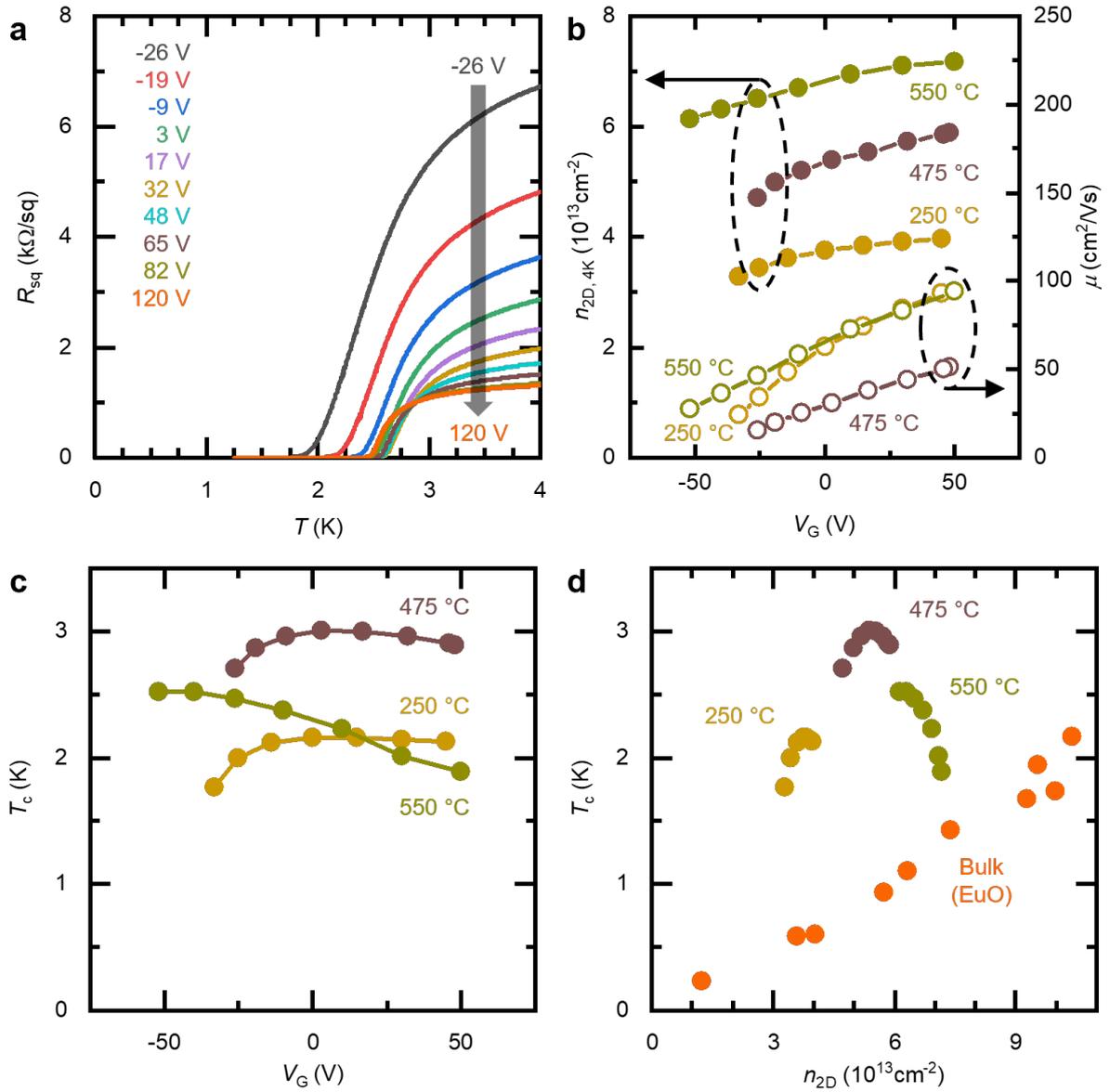

**Figure 3 | Gate-tunable superconductivity confirms dome behavior. a**, Electrostatic gating continuously tunes the superconducting transition. **b**, Gate voltage modulates carrier density and mobility. **c**, Non-monotonic $T_c$ variation with gate voltage. **d**, Comprehensive phase diagram combining growth and gating experiments reveals sharp superconducting dome. Quantum critical enhancement (up to 3 K) dramatically exceeds both undoped interfaces and bulk superconductivity. Bulk (EuO) is from C. Liu et al.[7]

# Supplementary Information

## Enhanced Superconductivity at Quantum-Critical KTaO$_3$ Interfaces

Jieun Kim[1,†], Muqing Yu[2,†], Ahmed Omran[2], Jiangfeng Yang[1], Ranjani Ramachandran[2], William O. Nachlas[3], Patrick Irvin[2], Jeremy Levy[2,*] and Chang-Beom Eom[1,*]

†These authors contributed equally to this work.
*Corresponding author(s). E-mail(s): jlevy@pitt.edu; ceom@wisc.edu;

**This PDF file includes:**

    Supplementary Figures 1-4

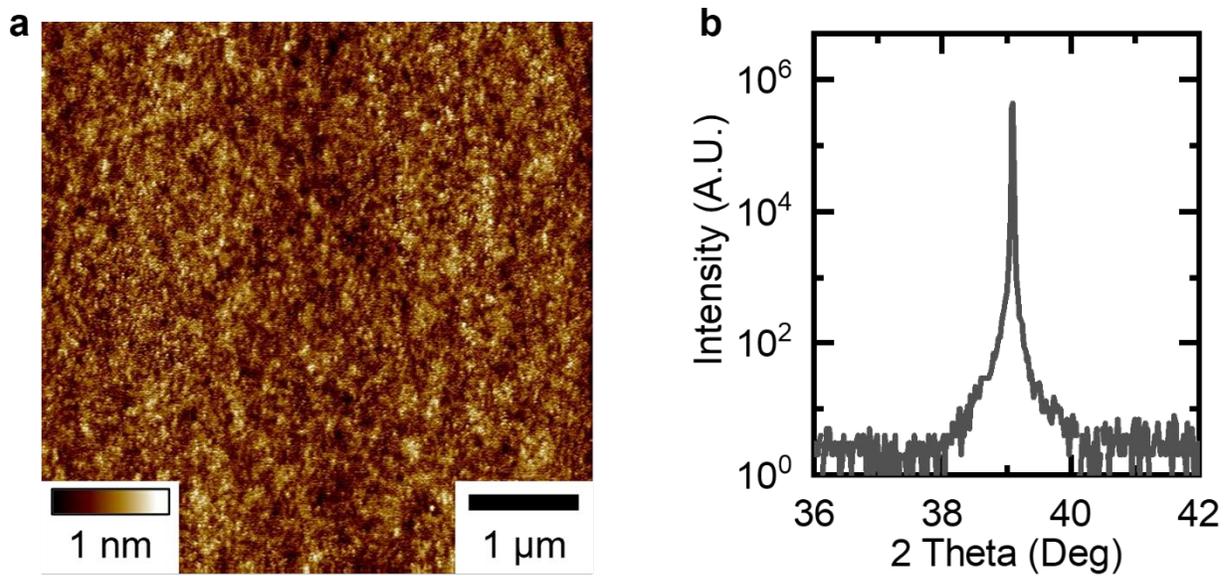

**Fig. S1. Structural characterizations of KTa$_{0.992}$Nb$_{0.008}$O$_3$/KTaO$_3$ (111) heterostructures. a**, Surface topography of ≈10 nm KTa$_{0.992}$Nb$_{0.008}$O$_3$/KTaO$_3$ (0111) heterostructures. **b**, θ-2θ X-ray diffraction line scans of ≈10 nm KTa$_{0.992}$Nb$_{0.008}$O$_3$/KTaO$_3$ (0111) heterostructures

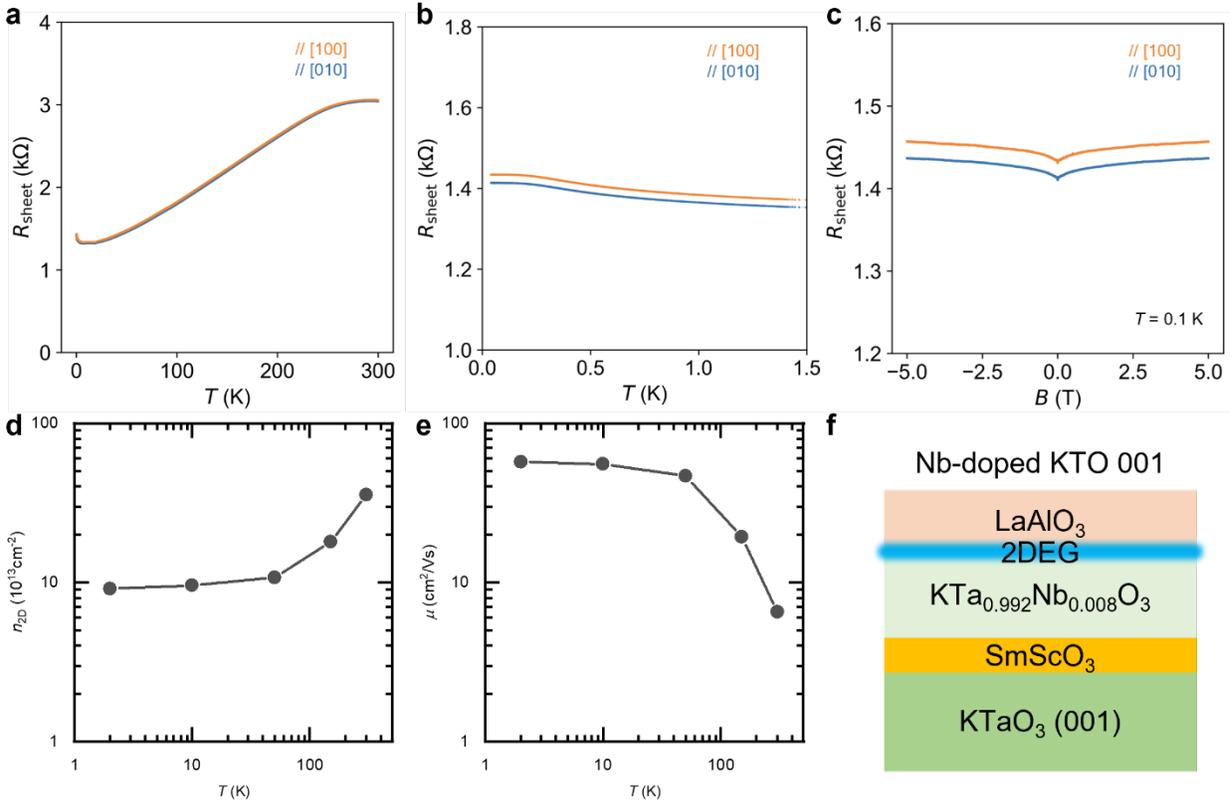

**Fig. S2. Electrical transport properties of LaAlO$_3$ / KTa$_{0.992}$Nb$_{0.008}$O$_3$ (001) / SmScO$_3$ / KTaO$_3$ (001) heterostructures showing the absence of superconductivity down to ~50 mK. a**, Temperature dependence of sheet resistance ($R_{sheet}$) from 300 to 2 K. **b**, Low-temperature $R_{sheet}$ down to ~50 mK showing no superconducting transition. **c**, Magnetoresistance at $T$ = 0.1 K up to 5 T, revealing no superconducting transition. **d**, Temperature dependence of carrier density ($n_{2D}$). **e**, Carrier mobility (μ) versus temperature. **f**, Schematic of the heterostructure.

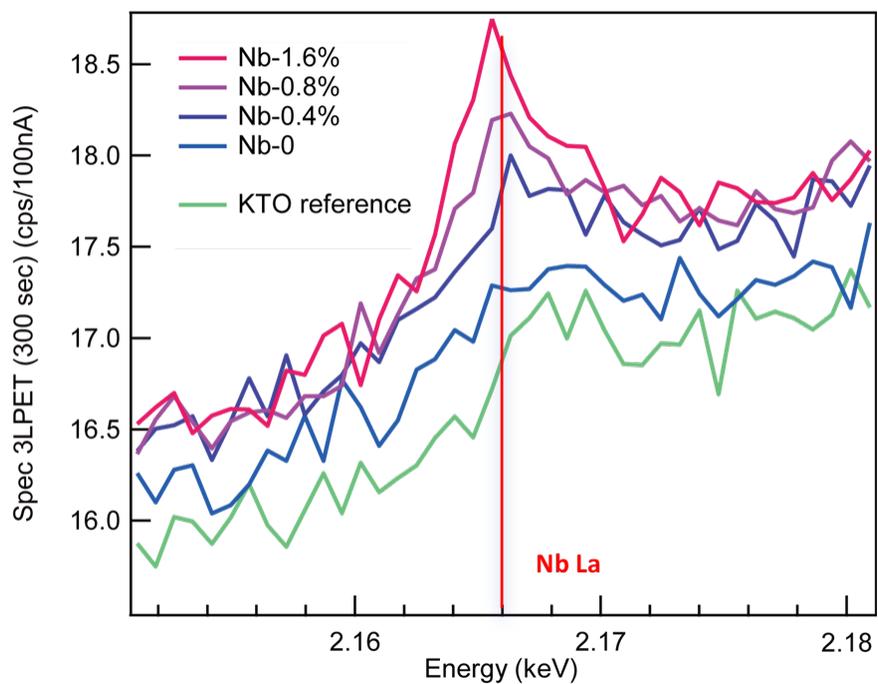

**Fig. S3. Wavelength-dispersive spectroscopy (WDS) spectra of KTaO$_3$ reference crystal and Nb-doped KTaO$_3$ films (0, 0.4, 0.8, and 1.6% Nb)**

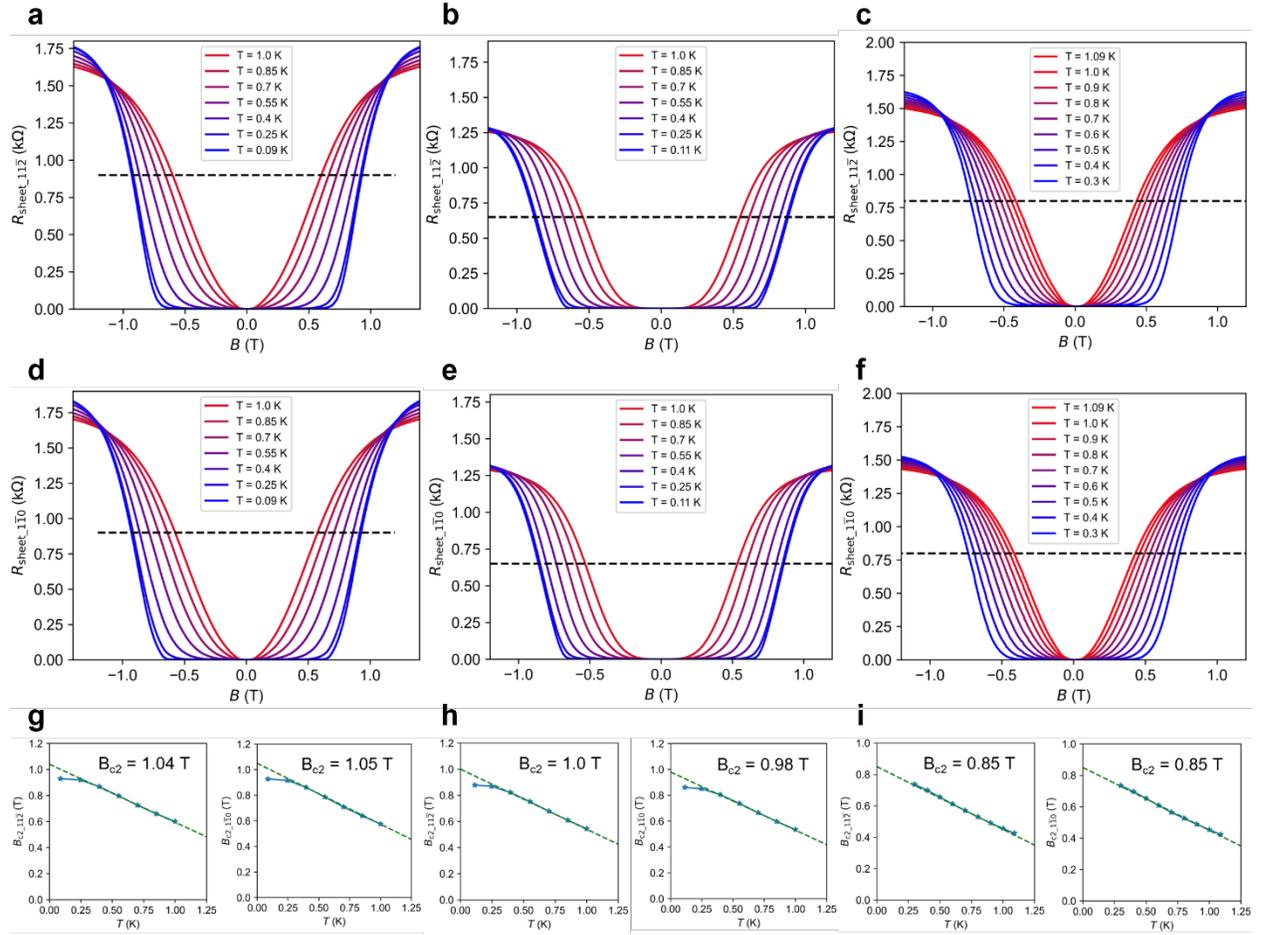

**Fig. S4. Temperature-dependent magnetoresistance and extracted upper critical fields ($B_{c2}$) for Nb-doped KTaO$_3$ (111) two-dimensional superconductors. a-c**, Sheet resistance ($R_{sheet}$) versus perpendicular magnetic field for devices patterned along the [11-2] direction with Nb doping levels of 0.4% (a), 0.8% (b), and 1.6% (c), measured at temperatures from 0.09 to 1.09 K. **d-f**, Corresponding $R_{sheet}$ vs. B curves for Hall bars along the [1-10] direction with Nb doping levels of 0.4% (d), 0.8% (e), and 1.6% (f), measured at temperatures from 0.09 to 1.09 K. In a-f, dashed horizontal lines mark the 50% normal-state resistance criterion used to extract $B_{c2}$. **g-i**, Temperature dependence of $B_{c2}$ for each device: 0.4% Nb along the [11-2] and [1-10] (g), 0.8% Nb along the [11-2] and [1-10] (h), and 1.6% Nb along the [11-2] and [1-10] (i).

5